\documentclass[sigplan,screen]{acmart}
\usepackage[ruled,vlined]{algorithm2e}
\usepackage{multirow}

\usepackage{booktabs}
\usepackage{subfig}
\AtBeginDocument{%
  \providecommand\BibTeX{{%
    \normalfont B\kern-0.5em{\scshape i\kern-0.25em b}\kern-0.8em\TeX}}}

\setcopyright{acmcopyright}
\copyrightyear{2021}
\acmYear{2021}
\acmDOI{ }

\acmConference[Nashville '21]{Nashville '21: ACM/IEEE International Conference on Cyber-Physical Systems}{May 19--21, 2021}{Nashville, TN}
\acmBooktitle{Nashville '21: ACM/IEEE International Conference on Cyber-Physical Systems,
  May 19--21, 2021, Nashville, TN}
\acmPrice{}
\acmISBN{}

\begin{document}

\title{Multimodal Mobility Systems: Joint Optimization of Transit Network Design and Pricing}


\author{Qi Luo}
\affiliation{%
  \institution{Clemson University}
  \city{Clemson}
  \state{SC}
  \country{USA}}
 \orcid{0000-0002-4103-7112}
\email{qluo2@clemson.edu}

\author{Samitha Samaranayake}
\affiliation{%
 \institution{Cornell University}
 \streetaddress{}
 \city{Ithaca}
 \state{NY}
 \country{USA}}
\email{samitha@cornell.edu}

\author{Siddhartha Banerjee}
\affiliation{%
  \institution{Cornell University}
  \city{Ithaca}
  \state{NY}
  \country{USA}}
 \email{sbanerjee@cornell.edu}

\renewcommand{\shortauthors}{Luo, et al.}

\begin{abstract}
The performance of multimodal mobility systems relies on the seamless integration of conventional mass transit services and the advent of Mobility-on-Demand (MoD) services. 
Prior work is limited to individually improving various transport networks' operations or linking a new mode to an existing system. In this work, we attempt to solve transit network design and pricing problems of multimodal mobility systems en masse.
An operator (public transit agency or private transit operator) determines  frequency settings of the mass transit system, flows of the MoD service, and prices for each trip to optimize the overall welfare.
A primal-dual approach, inspired by the market design literature, yields a compact mixed integer linear programming (MILP) formulation. However, a key computational challenge remains in allocating an exponential number of hybrid modes accessible to travelers. We provide a tractable solution approach through a decomposition scheme and approximation algorithm that accelerates the computation and enables optimization of large-scale problem instances. 
Using a case study in Nashville, Tennessee, we demonstrate the value of the proposed model.  
We also show that our algorithm reduces the average runtime by 60\% compared to advanced MILP solvers. 
This result seeks to establish a generic and simple-to-implement  way of revamping and redesigning regional mobility systems in order to meet the increase in travel demand and integrate traditional fixed-line mass transit systems with new demand-responsive services.

\end{abstract}




\maketitle

\section{Introduction} \label{sec:1}
Emerging Mobility-on-Demand (MoD) solutions, from ride-hailing platforms (e.g., Uber, Lyft) to on-demand bus services~\cite{alonso2017demand}, provide responsive and reliable mobility options to urban commuters. Nevertheless, there are growing concerns that MoD would compete with or substitute for conventional transit, increase traffic congestion, or adopt discriminatory price mechanisms \cite{edelman2015efficiencies,henao2019impact,luo2019dynamic}. This work proposes a new framework for building a multimodal mobility system as a first step in the full integration of Mobility-on-Demand (MoD) services and regional mass transport networks.  Mass transit (MT) 
operates as an affordable top-down system where users have to adapt to available routes and service times. On the other hand, MoD platforms  provide flexible door-to-door service but are not affordable for every traveler. This work aims to explore the opportunity to get the best of both systems by tightly integrating the two systems at the system design level, thereby understanding the true potential of a fully integrated multimodal mobility system. 

A primary motive for establishing a multimodal mobility system is to ensure that MT and MoD services are seamlessly integrated into an efficient transportation network. The system's success hinges upon the confluence of two modes and meeting the latent travel demand given user preferences. MT provides low-cost services along fixed routes to a prearranged timetable. In contrast, MoD can provide access to underserved communities by MT (due to various reasons) at a higher expense.  On a network design level, integrating MoD helps extend MT's reach through first- or last-mile connections; On an operational level, redirecting travelers by MoD can harness underused MT routes and reduce the overall operational costs.  This work aims to provide a tractable optimization framework that simultaneously design networks and determines service prices in favor of the public good.

\subsection{ Backgrounds } \label{sec:1-1}
Multimodal mobility systems are an emerging service that cuts across traditional transit boundaries and emerging MoD applications. Transportation agencies are playing a prominent role in shaping such services by experimenting with new service models and rules for governing integration with on-demand mobility service providers~\cite{pinto2019joint}. For example, the Chicago Regional Transportation Authority and Lyft launched an incentive program that connects commuters from nearby metro-stations to office buildings \cite{chicago2019}. The local transportation agency is responsible for allocating resources between improving MT services and subsidizing MoD services as first- or last-mile connections.  MoD, either owned by the public sector or a private company, can benefit from reaching a broader basis of customers  \cite{banerjee2020market}.  

A vast stream of literature has studied the \emph{centralized} planning, tactical, and operational decisions in transit or MoD systems. Planning decisions include choosing service regions or bus stop locations. Tactical decisions include setting bus routes' frequency. Operational decisions include timetabling of MT and routing of MoD.   Interested readers are referred to the reviews on MT \cite{ibarra2015planning} and this stream of work on MoD \cite{wang2019ridesourcing}.  More recently, joint decision-making on different levels has been studied thoroughly taking both short-term and long-term travelers' behavior into consideration.   Zhang et al. \cite{zhang2018public} investigated the joint optimization of MT service frequency and the fare to maximize a weighted sum of profit and consumers' surplus. Sun and Szeto \cite{sun2019optimal} evaluated the effectiveness of sectional fares in MT by solving a bilevel program that jointly determines the fares and frequency setting. These approaches do not apply to a multimodal setting due to the exponential blowup in computation time when including multiple modes. 

Modeling travelers' mode choice is a critical technical challenge of optimizing multimodal networks. The latent travel demand is realized only when the prices and the quality of service (QoS) are both satisfying. In general, this choice model is treated as a lower-level optimization in centralized decisions.   Beheshtian et al.  \cite{beheshtian2020bringing} studied the multimodal marketplace in which service providers bids for the use of road segments and travelers pay the roadway tolls by the clearing prices.  In addition,  transportation agencies may incentivize travelers to use one mode of transport over the others. Brands et al. \cite{brands2020tradable} designed a tradable permit scheme to regulate road transport externalities and empirically tested the existence of dynamic equilibrium on the mobility market.  Characterizing those user equilibria is a computationally heavy task, regardless of the recourse under various tactical and operational plans. A simplified framework is proposed by Wischik \cite{wischik2018price}. It studied the multimodal mobility market's knock-on effect on travelers' route and mode choices by combining a discrete choice model with a multipath resource allocation model.  By solving a single optimization problem,  a transportation agency can infer the choice models and then control traffic flows externalized by travelers' prices. Leveraging this idea,  this work focuses on solving a joint pricing and network design problem for multimodal mobility systems. 

\subsection{Contributions}  \label{sec:1-2}  To the best of our knowledge, this work is the first attempt to propose a scalable framework for solving the joint pricing and network design problem in multimodal mobility networks using exact methods.
This problem is notoriously difficult to solve even in single-mode transportation networks, and exacerbated by the complexities introduced in the multimodal setting. Two critical challenges in the multimodal setting are the growth of feasible combinations of modes and modeling user preferences. Our work computes the optimal prices for hybrid trips by a primal-dual approach developed in the market design literature. The computed prices ensure that each traveler is willing to pay for their utility-maximizing option. The optimal design of the marketplace achieves maximal social welfare.  
Besides, the multimodal mobility network needs to determine the frequency of MT routes and relocate MoD vehicle fleets to meet travel demand.  Since the number of possible modes is enormous, we combine approximation algorithms, Benders decomposition, and cutting-plane approaches into a unified solver to obtain implementable algorithm for optimal pricing and operational planning. This approach's performance in facilitating the convergence of the two modes is experimentally evaluated using a real-world case study in Nashville, Tennessee. 

The outline of this work is as follows. Section \ref{sec:2} describes the formulation of the joint pricing and network design problem.  Section \ref{sec:3} presents a decomposition scheme that takes advantage of the special structure of the subproblems.  Section \ref{sec:4} conducts a case study applying the developed approach to the city of Nashville. Finally, we draw some conclusions in Section \ref{sec:5}. 

\section{ Model}  \label{sec:2}

We study a joint pricing  and network design problem in a multimodal mobility system. With a welfare maximization goal in its mind, a transportation agency is responsible 
for pricing multimodal mobility services and  determining the routes and frequency setting of MT  to meet demand for affordable and efficient mobility services.  

\subsection{A motivating example for multimodal systems}  \label{sec:2-1}

\begin{figure}[!htb]
    \centering
    \includegraphics[width = 0.48\textwidth]{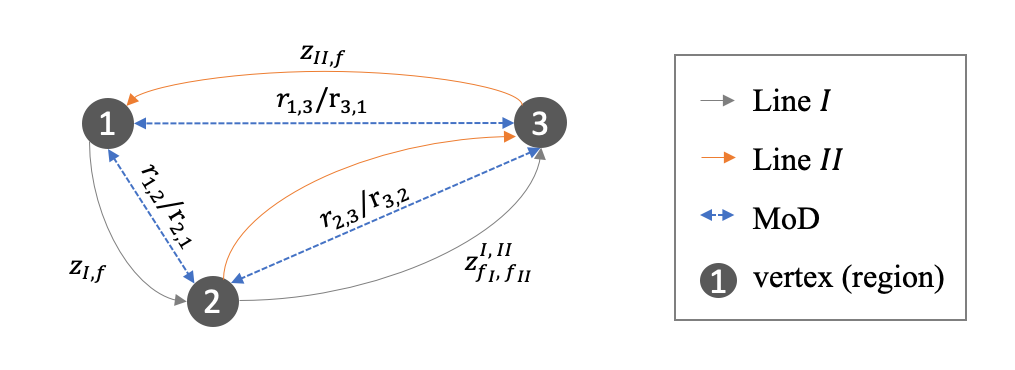}
    \caption{Pricing and network design problem in a simple multimodal network}
    \label{fig:1}
\end{figure}

The following  example (Figure \ref{fig:1}) shows how to coordinate the operations of  MT and MoD subject to travelers' behavior to maximize overall welfare. 

The network has two one-way MT lines, $I$ and $II$, and  MoD is available on each edge.   As MoD serves trip requests in an on-demand fashion, it needs to rebalance idle vehicles to high demand locations~\cite{pavone2012robotic}. The rebalancing flows between each pair of vertices $i,j$ at steady-state are $r_{ij}$.  There are two types of travelers $\theta_1$ and $\theta_2$ with different utility functions.  Travel demand from vertex $1$ to vertex $3$ is split to  $\lambda_{1,3}(\theta_1)$ and $\lambda_{1,3}(\theta_2)$, respectively.   Travelers of type $\theta_1$ are price-sensitive and travelers of type $\theta_2$ are QoS-sensitive. 

\begin{figure}[!htb]
	\centering
	\includegraphics[width = 0.48\textwidth]{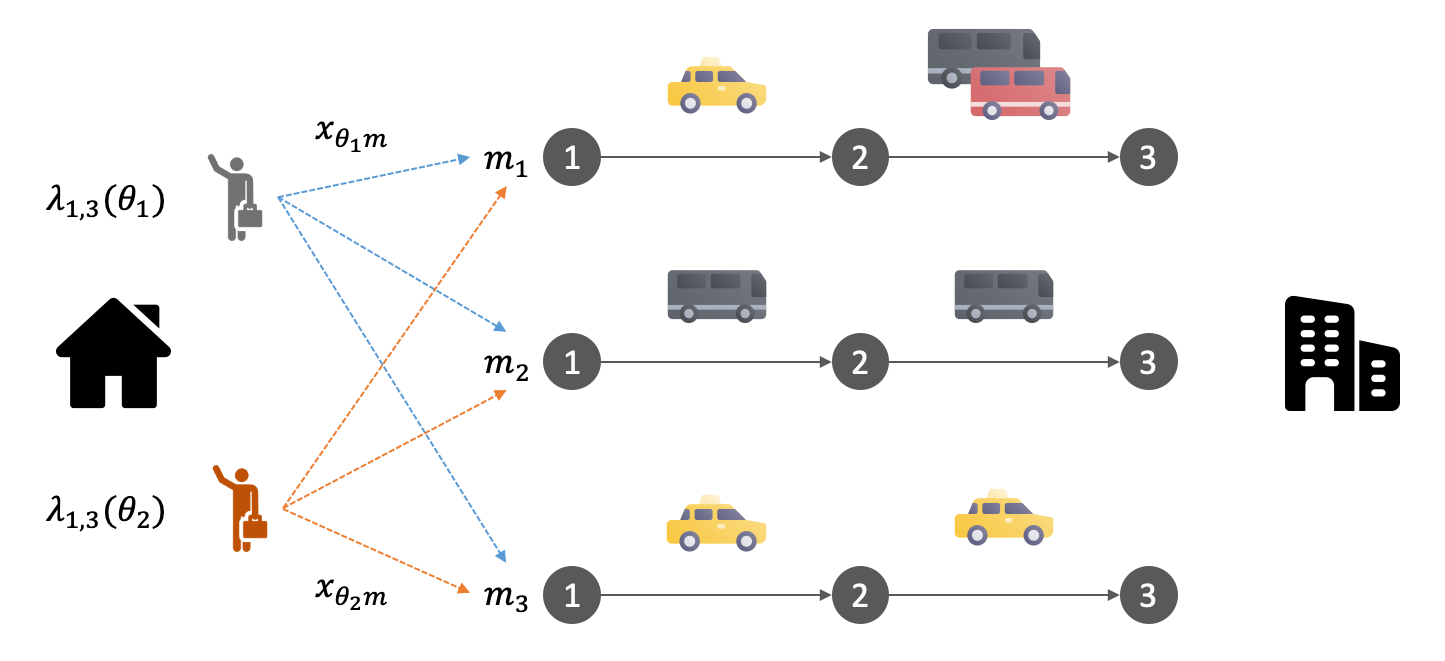}
	\caption{Travelers' choice model in a multimodal network}
	\label{fig:2}
\end{figure}

Travelers have unified access to use a mix of two mobility services through a trip planner.  For example, each traveler from vertex $1$ to vertex $3$ are provided with three options (called ``mode'' throughout this work): (a) taking low-frequency MT Line I, 
(b) taking MoD, or (c) a multimodal mode, i.e., taking MoD to vertex $2$ and transfer to MT Line $I$ or Line $II$, which collectively arrive at a high frequency (Figure \ref{fig:2}). We define the time between MT vehicles past a given point as the headway, and the
MT service frequency in unit of the average number of vehicles per unit time  is the inverse of the headway. 
Assume that Line $I$ and Line $II$ operate at frequencies of $f_I$ and $f_{II}$, respectively.   If a traveler chooses mode (c), she will take the first arriving  MT vehicle at the vertex $2$. We model this option by creating a bundle of two lines from which each traveler chooses  Line $i \in \{I, II\}$ by a probability of $f_i / (f_I + f_{II})$.    Since mode (b) is expensive and mode (a) may not meet travelers' QoS requirements, improving the competitiveness of mode (c) when appropriate is the primary goal of this work.

After the system presents the aforementioned modes and their corresponding prices and QoS,  each traveler chooses the mode $m$  that maximizes the $(v_m - p_m), m \in \{m_1, m_2, m_3\}$, where $v_m$ is the valuation and $p_m$ is the charged price (including both MT fares and MoD fares).   Without integrated multimodal access, $\theta_1$ travelers tend to take Line I and $\theta_2$ travelers tend to choose MoD. As a result, Line I is overloaded, and the MoD vehicle fleet stagnates at vertex $3$. 



From the transportation agency's perspective, the goal is to create a multimodal market in which the market clearance prices for all modes, including these hybrid modes, lead to maximal social welfare.  
This work focuses on frequency setting and rebalancing decisions in addition to pricing is because frequency setting is the most fundamental tactical decision in MT and rebalancing is the unique and central task in MoD operations. 
In this market design procedure, an obvious trade-off is the resource allocation between MT and MoD.  Providing high-frequency MT services on Line $I$ enhances the QoS while increasing its operating costs. In this example, it is potentially beneficial to subsidize some travelers to use the multimodal  mode (c)  to gain access to the low-cost Line $II$.   On the other hand, they should  not be over-subsidized to choose the MoD mode (b).     Since the system's welfare is measured by the difference between utilities received from serving travel demand and the incurred operating costs, this approach can redistribute realized traffic streams to improve MT's utilization while reducing the cost of operating the MoD vehicle fleet. 



\subsection{Setting}  \label{sec:2-2}

In a general setting, the multimodal mobility network is represented by a directed graph $G = (V, E)$ consisting of $n$ vertices corresponding to pickup/dropoff locations. There are $L$ transit lines where each line is a sequence of edges  $\{ e_{ij} \}$, which joins a sequence of distinct vertices. MoD is assumed to be available on each edge.    A traveler from the starting vertex $s$ to the terminal vertex $t$ is provided with a menu of available hybrid modes consisting of MT and MoD denoted as $\mathcal{M}_{st}$. She employs a mixed strategy,  i.e., the trip planner provides her a randomized option from the chosen modes to maximize her expected utility. Let $[K] := \{1, 2, \dots, K\}$ throughout this work and $[K]^2$ denote the Cartesian product of sets.  The notation used throughout this work is summarized in the nomenclature Table \ref{table:4} in Appendix \ref{app:1}.


\subsubsection{ Frequency setting in bundled service}

The service frequency on each line is a vector $\pmb{f} \in \mathbb{Z}^{|L|}$. There exists a natural lower bound such that, when setting the frequency at this bound, MT becomes noncompetitive for all travelers compared with other options. We rescale the frequency such $f_{\ell} \in [F]$ for all $\ell \in [L]$. The transit network's configuration is fixed throughout the frequency setting decision.  Let $z_{\ell, f}$ be a binary variable corresponding to line $\ell$ being set to frequency $f$, i.e., $z_{\ell f} = 1$ if the frequency $f$ is turned \emph{on} and $z_{\ell f} = 0$ if the frequency is \emph{off}. Line $\ell$'s vehicle capacity is $\mathcal{V}_{\ell}$.  The unit setup cost incurred for raising the frequency of a transit line is $o_\ell$ so the total setup cost of MT is $\sum_{ \substack{\ell \in [L],  f\in [F]  }  }   o_{\ell} z_{\ell, f} $. 

When a set of MT lines $L_{ij}$  passes through the same edge $e_{ij}$, the system creates a set of bundled services  $\mathcal{B}_{ij} :=  \text{comb}(L_{ij})$. Each bundled service $b_{ij} \in \mathcal{B}_{ij}$ includes one or more MT lines.  The probability of a traveler who chooses  $b_{ij} $ taking line $\ell \in b_{ij}$ is proportional to its frequency  $ f_{\ell} /  \sum_{\ell' \in b_{ij} } f_{\ell '}$, i.e., on a first-come-first-serve (FCFS) basis.   The on/off decision for each bundle is denoted a binary variable $z_{\psi}^{b_{ij}}$ where $\psi_{\ell}$ is the corresponding frequency in $\psi$ at its $\ell \in b_{ij}$ entry. In example of Figure \ref{fig:1}, $b_{23} = \{I, II\}$ and $\psi = \{f_{I}, f_{II}\}$. $z_{\psi}^{b_{ij}} = 1$ means that the system provides bundled service $b_{ij}$ to all modes passing the edge $e_{ij}$. As each traveler is allowed to transfer at most once, a preprocess truncates the set of modes $\mathcal{M}_{st}$ when it is expanded to include bundled services at varying frequencies. 


\subsubsection{Demand model}
We consider a large market, fluid scaling of the system. Each vertex $s$ has exogenous, non-atomic flows $\{\lambda_{st}, t\in V\slash \{s\} \}$. Heterogeneous travelers of type $\theta \in \Theta$ have valuations drawn from a discrete distribution $q$. Thus the flows are split as $\lambda_{st}(\theta) := \lambda_{st} q_{st}(\theta)$.  Travelers from a starting vertex $s$ to a terminal vertex $t$ plan their trips from options (called ``hybrid  modes'') $m \in \mathcal{M}_{st}$ defined by a sequence of edges, each specifying the mode of transport as $\{(s, i_1), (i_1, i_2), ... (i_k, t)\}$. 

Observing existing services and their corresponding QoS and prices, travelers  make their trip plans (possibly using a multimodal option).  A traveler from vertex $s$ to vertex $t$ chooses the mode $m \in \mathcal{M}_{st}$ that maximizes their utility $v_{\theta m} - p_{\theta m}$. The valuation for trip $(s,t)$ is type- and mode-dependent, and denoted as $\pmb{v}(\theta) \in \mathbb{R}^{|M_{st}|}$.  $x_{\theta m} \in  [0, 1]$ denotes  the probability that a traveler of type $\theta$ traveling from $s$ to $t$ chooses mode $m \in \mathcal{M}_{s t}$ and $\sum_{m \in  \mathcal{M}_{s t} } x_{\theta m} = 1$. 

Since MoD and MT have different pricing practices, their operational costs are calculated separately in each hybrid mode $m \in \mathcal{M}_{st}$.  The total operational cost of hybrid
mode $m$ is given by $c_m =  \sum_{e_{ij} \in E_m} c_{ij} + \sum_{\ell \in L: \ell \sim m }  c_{\ell}$, where $\ell \sim m$ means that mode $m$ uses Line $\ell$.
Payments for MoD are distance-based, and we let $c_{ij}$ denotes the operating cost of carrying passengers by  MoD from $i$ and $j$. Payments for MT are made per-use.  The chosen Line $\ell$'s  fares depend on the od pair $s-t$ and the operational cost of Line $\ell $ is $c_{\ell}$.


\subsubsection{MILP formulation}
This section proposes a generic MILP formulation for the joint pricing and network design problem in \eqref{eq3}. The transit network design problem determines the MT frequency $f$ for all $\ell \in [L]$ ($\pmb{z}$) and rebalancing flows of MoD ($\pmb{r}$).  The objective is maximizing the overall welfare collected from operating the multimodal mobility system.  The welfare measures the difference between the utility of serving all travel demand, operational costs of hybrid modes, MoD's rebalancing costs , and MT's setup costs. 

The constraints are categorized as follows:
\begin{enumerate}
	\item   \eqref{eq3a} are the capacity constraints on each edge. This constraint guarantees that the total MT flow passing through edge $e_{ij}$ (left-hand side) is no larger than the total capacity carried by MT lines (right-hand side). The flow distribution for each  MT line is satisfied by enforcing a FCFS rule in trip assignment.
	\item \eqref{eq3b} - \eqref{eq3c} describe the relationship between MT's line setups and bundled service setups. A bundle $b_{ij}$ is accessible if and only if all lines included $\ell \in b_{ij}$ are operated at the frequency levels of $\psi$.
	\item \eqref{eq3d} - \eqref{eq3e} are the ranking constraints for MT frequency setting.  When the system determines Line $\ell$'s frequency to be at level $\ell$, low-frequency options are turned on as travelers tend to choose a high-QoS service when paying the same price. 
	\item  \eqref{eq3f} is the budget constraint for MT.  The cycle time of Line $\ell$ is $T_{\ell}$ and the total number of MT vehicles is $\mathcal{R}$.
	\item   \eqref{eq3g} ensures the flow-balance (via  rebalancing) of the MoD fleet.
	\item  \eqref{eq3h} is a linkage constraint in mode generation. A mode $m$ is available if each $\ell \sim m$ is open. It can be relaxed to $C x_{\theta m} \leq z_{\ell, f}$ with arbitrary $C>0$. 
	\item \eqref{eq3i} are the demand conservation constraints.  
\end{enumerate}

\noindent  \begin{minipage}{\linewidth}
	\centering
	\begin{subequations} \label{eq3}
		\noindent 
		\begin{align}
			& \max_{ \pmb{z, x, r}}  \sum_{ \substack{ (s,t)\in [n]^2  \\ \theta \in \Theta_{st} }   }     \lambda_{st} (\theta)  \sum_{m\in \mathcal{M}_{st}}     	  (v_{\theta m}  - c_m) x_{\theta m }  - \sum_{i,j} c_{ij} r_{ij} \nonumber \\
			& \qquad - \sum_{ \substack{\ell \in [L] \\  f\in [F]  }  }   o_{\ell} z_{\ell, f}   \tag{\ref{eq3} } \\
			&  s.t.		\sum_{ \substack{  (s,t): e_{ij} \in (s, t) \\ \theta \in \Theta, m\in \mathcal{M}_{s t}   } }   \lambda_{st}(\theta)   x_{\theta m}  \leq 
			\sum_{ \substack{ (\ell, f) \in (b_{ij}, \psi),  \\ \forall  \psi \in [F]^{|b_{ij}|}, \\ \forall b_{ij} \in \mathcal{B}_{ij}   } }  \mathcal{V}_{\ell} z_{\ell, f},  \, \forall e_{ij} \in E  \label{eq3a} \\ 
			&  z^{b_{ij}}_{\psi} \leq z_{\ell, f},  \qquad  \forall (\ell, f) \in (b_{ij}, \psi),  \forall  \psi \in [F]^{|b_{ij}|}, \nonumber \\
			& \qquad  \qquad \qquad \,  \forall b_{ij} \in \mathcal{B}_{ij}, \forall e_{ij} \in E 	\label{eq3b} \\
			& z^{b_{ij}}_{\psi } \geq \sum_{(\ell, f) \in (b_{ij}, \psi)  } z_{\ell, f} - |b_{ij}| + 1,  \qquad  \forall \psi\in [F]^{|b_{ij}|}, \nonumber \\
			& \qquad \qquad \qquad  \forall b_{ij} \in \mathcal{B}_{ij}, \forall e_{ij} \in E \label{eq3c} \\
			&  z_{\ell, f} \leq z_{\ell, f'},  \qquad \forall  f \geq f', \forall \ell \in [L] \label{eq3d} \\
			&  z_{\ell, f} \geq z_{\ell, f''},  \qquad \forall f \leq f'', \forall \ell \in [L]  \label{eq3e}\\ 
			& \sum_{\ell \in [L]}  \sum_{f \in [F]} T_{\ell } z_{\ell, f} \leq \mathcal{R} \label{eq3f} \\
			&    \sum_{j \in [n] } \left( r_{ji} + \sum_{  \substack{ \theta \in \Theta  \\ m: (j,i)\in E_m  }   }  \lambda_{st}(\theta)  x_{\theta m} \right)  =  \nonumber \\
			&  \sum_{k \in [n]} \left( r_{i k} + \sum_{ \substack{\theta \in \Theta \\ m: (i,k)\in E_m   }  }    \lambda_{st}(\theta) x_{\theta m} \right),
				\qquad  \forall i \in [n]  \label{eq3g} \\
			& x_{\theta m}  \leq z_{\ell, f}, \quad  \forall (\ell,f) \sim m, \forall \theta \in \Theta, \forall \ell \in {L}, \forall f \in [F]  \label{eq3h} \\
			& \sum_{ m \in \mathcal{M}_{st}^M } 	x_{\theta m} \leq q_{st}(\theta)	, \qquad  \forall \theta \in \Theta, \forall (s,t) \in [n]^2  \label{eq3i} \\
			& z_\psi^{b_{ij}} \in \{ 0,1 \},  \qquad \forall \psi \in [F]^{ |b_{ij}| }, \forall b_{ij} \in \mathcal{B}_{ij}, \forall e_{ij } \in E  \label{eq3j} \\
			& z_{\ell,f} \in \{0,1\} ,  \qquad \forall f \in [F], \forall \ell \in [L] \label{eq3k} \\
			&  0 \leq x_{\theta m} \leq 1, \qquad  \forall \theta \in \Theta, m \in \mathcal{M}_{st}, (s, t) \in [n]^2 \label{eq3l} \\
			& r_{ij} \geq 0, \qquad \qquad \forall e_{ij} \in E  \label{eq3m}
		\end{align}
	\end{subequations}
\end{minipage}

\smallskip 

This formulation overcomes two main technical challenges in prior work. First, the transit network's design subject to travelers' mode choice is modeled by a primal-dual approach. This approach conserves the user equilibrium and avoids the computational tractability issue in the conventional bilevel programming approach \cite{sun2019optimal}.  Second,  considering $p_m$ as primal variables under the demand elasticity leads to a nonlinear mathematical program. We convert the pricing decisions into a dual problem (explained in Section \ref{sec:2-2-4}) and simplify the formulation by linearization tricks.  
This model can be extended to including the MoD platform as a separate decision-maker \cite{banerjee2020market}. The privately-owned MoD company is better off by participating in this multimodal network to gain a broader basis of customers.

\subsubsection{Dual prices for hybrid modes} \label{sec:2-2-4}

The LP-relaxation of \eqref{eq3} computes the dual prices $p_{\theta m}$ for each mode $m \in \mathcal{M}_{st}$.  The dual variables for the MILP formulations are as follows: 
\begin{enumerate}
	\item Edge-base price: $p_{ij}$ for each $e_{ij} \in E$ corresponding to \eqref{eq3a}.
	\item Bundled price: $\beta_{\psi}^{b_{ij} } $ and $\eta_{\psi}^{b_{ij} } $ for  $(b_{ij}, \psi)$ $ \in (\mathcal{B}_{ij}, |F|^{|b_{ij}|}  )$ and $e_{ij} \in E$ corresponding to \eqref{eq3b} and \eqref{eq3c}.
	\item Line price: $w_{\ell, f, f'}$ with $f\leq f', f,f'\in [F]$ corresponding to \eqref{eq3d} and \eqref{eq3e}.
	\item MT vehicle setup price: $u$ corresponding to \eqref{eq3f}.
	\item MoD connection price: $\gamma_i$ for $i \in [n]$ corresponding to \eqref{eq3g}.
	\item Line setup cost: $\zeta_{\theta m \ell f}$ for all $  (\ell,f) \sim m,  \theta \in \Theta,  \ell \in {L},  f \in [F] $ corresponding to \eqref{eq3h}.
	\item Path-base utility: $\nu_{\theta, st}$ for all $ \theta \in \Theta, m \in \mathcal{M}_{st}, (s, t) \in [n]^2$ corresponding to \eqref{eq3i}.
\end{enumerate}

Traveler of type $\theta$ traveling from $s$ to $t$ is charged a price $p_{\theta m}$ computed x  as follows:
\begin{align}
	p_{\theta m} &= \sum_{e_{ij} \in E_m} \left[ p_{ij}   +  \beta_{\psi}^{b_{ij} } +  \eta_{\psi}^{b_{ij} } +  \gamma_i 
	\right] + w_{\ell, f, f'} +  u  + \zeta_{\theta m \ell f},   \nonumber  \\
	& m \sim (b_{ij}, \psi),  
\end{align} 
where $m \sim (b_{ij}, \psi)$ means that the bundled service on edge $e_{ij}$ is included in the hybrid mode $m$; $E_m$ and $V_m$ are edges and vertices trespassed by this mode.  

It is worth mentioning that the dual prices are not well-defined in the original MILP  \eqref{eq3}.  Section \ref{sec:3} justifies the definition of dual prices with a decomposition scheme.  After fixing the frequency setting of MT, the travelers' mode choices are formulated as an LP \cite{wischik2018price} and the pricing  problem regarding the realized network flow is precisely defined. Solving these subproblems will return part of the dual price 
and feasibility/optimality cuts.  The algorithm then resolves the master problem to approach a feasible frequency-setting plan for MT such that the total welfare is maximized.  In the final step, extra setup costs for MTs are added evenly for all users, reflecting the equal-sharing practice in public transportation sectors.  This decomposition is also  computationally efficient as the subproblems are solvable in polynomial time.

\begin{proposition} \label{prop1}
	Under the optimal dual prices $p_{\theta}$, travelers of $\theta$ from vertex $s$ to  vertex $t$ in graph $G$   will adopt a mixed strategy $x_{\theta m}^*$ over available modes $m \in \mathcal{M}_{st}$ such that 
	\begin{equation}
		x_{\theta m} \in  \arg \max \left\{  \sum_{ \substack{  \theta \in \Theta, \\ m \in \mathcal{M}_{st}  } }  ( v_{ \theta m }- p_{\theta m}  ) x_{\theta m} \right\}, \, \forall (s,t) \in [n]^2.	\nonumber
	\end{equation}
\end{proposition}

This mixed strategy for choosing modes can be implemented by a randomized assignment policy as follows. Observing that each optimal mode $m$ gives type $\theta$ travelers the same utility, the system can randomly provide mode $m$ with a probability of $x^*_{\theta m}$ and keep track of the past assignment. In the following assignment, the system matches the empirical distribution of mode assignments with these $x^*_{\theta m}$.

\section{Pricing services with decomposition} \label{sec:3}
The joint optimization of transit network design and pricing formulated above is notoriously hard to solve directly regarding the large-scale urban networks. This section starts with a negative result for the integrality gap of the original MILP formulation. To overcome the computational challenge, we design a bilevel decomposition scheme such that the dual prices of modes are computed in recourse. Finally, we shed light on the computational complexity of the proposed algorithm.

\subsection{Overview of decomposition framework}
The dual prices can be computed from the constraints \eqref{eq3a} -\eqref{eq3f} in the LP-relaxation of \eqref{eq3}. Nevertheless, since both the frequency setting $z$ and mode choice $x$ are integer variables, we show that the integrality gap of the original formulation is arbitrarily large. 

In a simple network of two vertices, $L=1$, and $F$ equals a constant $\hat{F}$, we consider only MT in a round trip.  Without loss of generality, we assume the optimal frequency is $f^*$ such that $\sum_{f\in [f^*]} T_I z_{I,f} \leq \mathcal{R}$ and $\sum_{f\in [f^* + 1]} T_I z_{I,f} >  \mathcal{R} $. Hence, $z_{I, 1} = \dots = z_{I, f^*} = 1$ and $z_{I, f^*+1} = \dots z_{I, F} = 0$.  We can denote the mode choice variable $x_f$ in this case as the modes only differ in frequency.  The setting up cost is $ f^* o_{\ell} $ and the total capacity is $f^* V_I$. The constraint \eqref{eq3a} determines $x_{f}$ as $x_{f} > 0$ for the maximal $v_{\theta f} - c_f$ for $f \in [f^*]$ and equals $0$ for others. We denote the objective value as $v^*$.  We can find fractional solutions for its LP-relaxation counterpart as follows. We set $z_{I, f} =  f^* \slash \hat{F} $ for all $f \in [\hat{F}]$ which is feasible for the constraints in \eqref{eq3}. The total capacity and the setup costs are equal to the original problem while all modes are open. $x_f$ is expanded to $f\in [\hat{F}]$ and the integrality gap  $\max\{ v_{\theta f} - c_f\}_{  f\in [f^*+1, \hat{F}] } \slash v^*$ is unbounded. 

\begin{figure}[!htb]
	\centering
	\includegraphics[width = 0.48\textwidth]{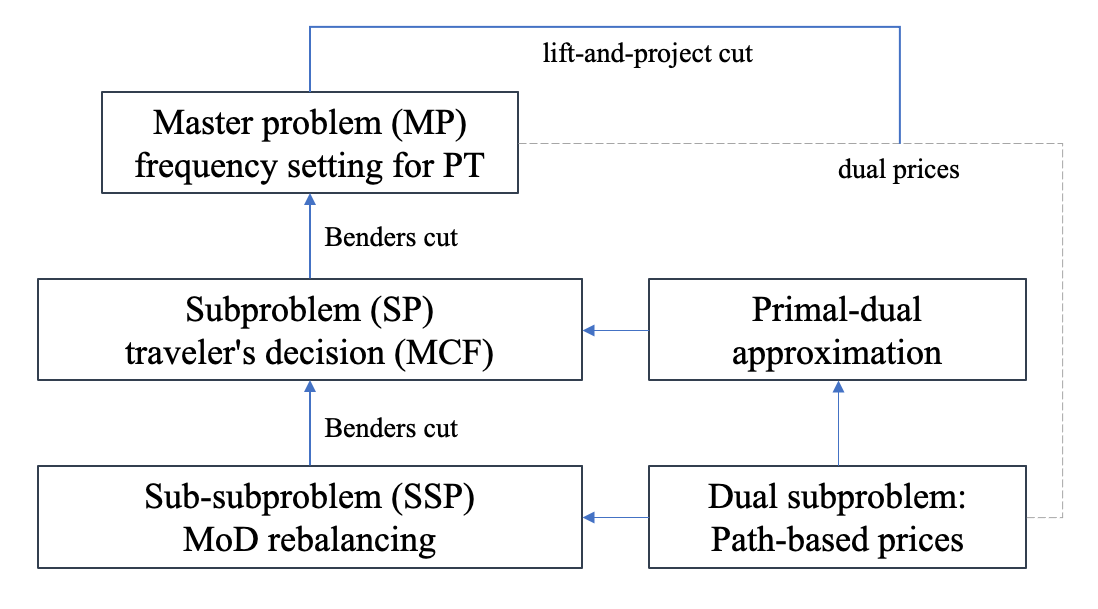}
	\caption{Decomposition framework for joint pricing and transit network design problem }
	\label{fig:3}
\end{figure}

The large integrality gap motivates the development of a decomposition framework in Figure \ref{fig:3}. The original MILP is computationally intractable as the dimension of binary variables $z$ and $x$ are in the scale of $2^{L \times F \times |E|}$ and the resulting number of constraints explodes. 


We decompose the MILP formulation to two levels of subproblems that solve $z$, $x$, and $r$ sequentially.

\subsubsection{Subproblem: traveler choice} 
We start with subproblems for  any transit  network design, which are solved through fully polynomial time approximation schemes. 
Let $y_{\theta m} = \lambda_{st}(\theta) x_{\theta m}$ and the values from the subproblem $\hat{y}_{\theta m} = \lambda_{st}(\theta) \hat{x}_{\theta m} $.  Given a MT network design  $\pmb{z}$ and  travelers' choice $\pmb{y}$,	the sub-subproblem $SUB(\pmb{z, y})$  (the MoD rebalancing problem)  is a min-cost flow problem:
	\begin{subequations} \label{eq6}
		\begin{align}
			& SUB(\pmb{z, y}) = 	\min_{ \pmb{ r}} \sum_{i,j} c_{ij} r_{ij}  \tag{\ref{eq6}  } \\
			s.t.  
			&  \sum_{j \in [n] } \left( r_{ji} + \sum_{  \substack{ \theta \in \Theta  \\ m: (j,i) \\ \in E_m  }   }   \hat{y}_{\theta m} \right)  = \nonumber  \sum_{k \in [n]} \left( r_{i k} + \sum_{ \substack{\theta \in \Theta \\ m: (i,k) \\ \in E_m   }  }     \hat{y}_{\theta m} \right), \\
			 & \qquad \qquad \qquad  \forall i \in [n]  \label{eq6b} \\
			& r_{ij} \geq 0, \qquad \forall e_{ij} \in E  \label{eq6f}
		\end{align}
\end{subequations}

The dual of the MoD rebalancing problem is 
	\begin{align} \label{eq7}
		& \max_{ \pmb{\gamma} } \sum_{i\in [n]} \left( \sum_{ \substack{\theta \in \Theta \\ m: (i,k)\in E_m   }  }    \hat{y}_{\theta m}  -  \sum_{  \substack{ \theta \in \Theta  \\ m: (j,i)\in E_m  }   }    \hat{y}_{\theta m} \right) \gamma_i \\
	 s.t. \, 	&  \gamma_i  - \gamma_j  \leq c_{ij}, \qquad \forall  e_{ij} \in E \nonumber \\
	 & \gamma_i  \, \text{ unrestricted},  \qquad \forall i \in [n] \nonumber
		\end{align}

The feasibile set of the dual problem is independent of $\pmb{x}$. Solving the subsubproblem adds either feasibile cuts or optimal cuts to the subproblem $P(\pmb{z})$ in \eqref{eq5}:
\begin{subequations} \label{eq5}
	\begin{align}
		&P(\pmb{z}) = 	\max_{ \pmb{y}} \sum_{ \substack{(s,t)\in [n]^2 \\ \theta \in \Theta_{st}  \\ m\in \mathcal{M}_{st}  }    }      	  (v_{\theta m}  - c_m) y_{\theta m }  + \phi
		\tag{\ref{eq5}  } \\
		& s.t. \quad  \eqref{eq3a}, \eqref{eq3h}, \eqref{eq3i}	 \\ 
		& \sum_{i \in [n]^e } \left( \sum_{ \substack{\theta \in \Theta \\ m: (i,k)\in E_m   }  }     y_{\theta m}  -  \sum_{  \substack{ \theta \in \Theta  \\ m: (j,i)\in E_m  }   }  y_{\theta m} \right) \gamma_i^e \geq \phi  \label{eq5b}\\
		&  \sum_{i \in [n]^u ]} \left( \sum_{ \substack{\theta \in \Theta \\ m: (i,k)\in E_m   }  }     y_{\theta m}  -  \sum_{  \substack{ \theta \in \Theta  \\ m: (j,i)\in E_m  }   }  y_{\theta m} \right) \gamma_i^u \geq 0 \label{eq5c} \\
		&   0\leq y_{\theta m} \leq \lambda_{st}, \qquad  \forall \theta \in \Theta, m \in \mathcal{M}_{st}, (s, t) \in [n]^2  \label{eq5d}
	\end{align}
where $\gamma^e$ are extreme points of \eqref{eq7} and $\gamma^u$ are unbounded directions of \eqref{eq7}.
\end{subequations}

The traveler choice subproblem \eqref{eq5} is a  variation of the weighted multicommodity flow problem (WMFP). If each traveler adopts a deterministic mode choice strategy, it is reduced  to an integral MFP problem, which is a well-known NP-complete problem (MAX SNP-hard). 
Compared to a direct LP approach, the subproblem computation's main bottleneck is the number of commodities $|\Theta|\cdot |\mathcal{M}_{st}|$  in multimodal mobility networks.  A primal-dual approximation algorithm (Algorithm \ref{alg:1}) is used to simultaneously compute travelers' routes and their corresponding prices. The essence of the algorithm is repeated solve shortest-path problems on the dual solutions to WMFP such that the approximation ratio is independent of the number of commodities \cite{fleischer2000approximating}. 

The dual problem of the fractional subproblem \eqref{eq5} is as follows:
\begin{subequations} \label{eq8}
	\begin{align}
		&D(\pmb{z}) =   \min  \sum_{ \substack{ (\ell, f) \in (b_{ij}, \psi),  \\ \forall  \psi \in [F]^{|b_{ij}|}, \\ \forall b_{ij} \in \mathcal{B}_{ij}   } } V_{\ell} \hat{z}_{\ell, f} p_{ij} +  \sum_{ \substack{  \theta \in \Theta \\ m \in \mathcal{M}_{st} \\ (\ell,f) \sim m  }  }  \hat{z}_{\ell, f} \zeta_{\theta m \ell f}   \nonumber \\
		& \quad  \sum_{\substack{\theta \in \Theta \\ (s,t) \in [n]^2 }} \lambda_{st}(\theta) (\lambda_{st}(\theta) \nu_{\theta, st} + \alpha_{\theta m} )  \tag{\ref{eq8}  }  \\
		s.t. & \, \sum_{e_{ij} \in E} p_{ij} + \sum_{(\ell, f)\sim m} \zeta_{\theta m \ell f} + \alpha_{\theta m }  + \nu_{\theta, st} \geq v_{\theta m}  - c_m,  \nonumber  \\
		& \qquad \forall \theta \in \Theta, \forall m\in \mathcal{M}_{st}, \forall(s,t) \in [n]^2  \label{eq8a} \\
		& A_{\phi}^{\intercal}  \pmb{\beta  } \geq 0    \\
		& \pmb{p}, \pmb{\zeta},\pmb{\nu}, \pmb{\alpha}, \pmb{\beta} \geq 0
	\end{align}
\end{subequations}
where $\pmb{\rho, \zeta, \nu, \alpha, \beta}$ are dual variables corresponding to constraints \eqref{eq3a}, \eqref{eq3h}, \eqref{eq3i}, \eqref{eq5d}, \eqref{eq5b} - \eqref{eq5c}, respectively. $A_{\phi}$ is the coefficient matrix of \eqref{eq5b}  and \eqref{eq5c}. We denote the objective function of the dual  problem as $g(\pmb{p}, \pmb{\zeta},\pmb{\nu}, \pmb{\alpha}, \pmb{ \beta }) $.

\begin{algorithm}[!htb]
	\SetAlgoLined
	\KwResult{Dual prices related to traveler choice and benders cuts \eqref{eq9b}, \eqref{eq9c}. }
	Initialization: frequency setting $\hat{z}$ in a network\;
	$\qquad$ $\pmb{y} = 0$, $\pmb{p}, \pmb{\zeta},\pmb{\nu}, \pmb{\alpha}, \pmb{\beta} = \delta$\;
	\While{new cuts are generated from \eqref{eq7}
	}{
	\For{$(s,t) \in [n]^2$,  $i \in [ \log_{1+\epsilon} \frac{1+\epsilon}{\delta }], j\in [K] $ }{
	$P \leftarrow \min_{D(\hat{z})} \mathcal{P}_j$\;
		\While{$p(\mathcal{P}_i^{st}) < \min \{1, \delta(1+\epsilon)^i \}$ }{
		$V^* \leftarrow \min_{e_{ij}\in P} V_{\ell} \hat{z}_{\ell,f} $ \;
		$y_{\theta m} \leftarrow y_{\theta m} + V^*$\;
		$\forall e_{ij} \in P$, $p_{ij} \leftarrow p_{ij} (1+\frac{\epsilon V^*}{V_{\ell} \hat{z}_{\ell,f}  }) $
		Update $P$.
	}
	}
	
	Solving subsubproblem $SUB(\pmb{z}, \pmb{y})$ with $\hat{\pmb{y}}$ and add either feasibility or optimality cuts.
	}
	\caption{Approximation subroutine for traveler choice problem} \label{alg:1}
\end{algorithm}

The approximation algorithm for WMCF (Algorithm \ref{alg:1})
provides a $(1-\epsilon)$ approximation, where $\epsilon$ is the error tolerance. The number of constraints in \eqref{eq5} is upper-bounded by $ \bar{m} := n^2  |\Theta|  L F$.  Let $\delta = \epsilon / \bar{m}^2$ be the initial value of dual variables in \eqref{eq8}. The number of commodities is $K = |\mathcal{M}_{st}|\cdot |\Theta|$.  The paths of commodity $i$ for each $(s,t)$ is denoted as $\mathcal{P}^{st}_{i}$.    We assume that, using the dual values in \eqref{eq8}, there exists a shortest-path oracle that finds the path with minimum cost for  $\mathcal{P}^{st}_{i}$ in polynomial time as $\min_{D(\hat{z})} \mathcal{P}_i^{st}$. The shortest path computed using the dual values is $p(\mathcal{P}_i^{st})$. Dual prices for each commodity are derivatives from the subroutine in  Algorithm \ref{alg:1}.


\subsubsection{Master problem: frequency setting} 
The top level of the original formulation \eqref{eq3} is a frequency setting problem with bundled services:
\begin{subequations} \label{eq9}
	\begin{align}
			& 	\max_{ \pmb{z}}  \tau -   \sum_{ \substack{\ell \in [L]  \\ f\in [F]   } }   o_{\ell} z_{\ell, f}   \tag{\ref{eq9}  } \\
			s.t. & \, \eqref{eq3b} - \eqref{eq3f}, \eqref{eq3j} - \eqref{eq3k}  \\
			& g(\pmb{p}^e, \pmb{\zeta}^e,\pmb{\nu}^e, \pmb{\alpha}^e, \pmb{\beta}^e )    \geq \tau  \label{eq9b}\\
			& g(\pmb{p}^u, \pmb{\zeta}^u,\pmb{\nu}^u, \pmb{\alpha}^u, \pmb{\beta}^u ) \geq 0  \label{eq9c}
		\end{align}
\end{subequations}
$(\pmb{p}^e, \pmb{\zeta}^e,\pmb{\nu}^e, \pmb{\alpha}^e, \pmb{\beta}^e) $ are extreme points of the dual subproblem \eqref{eq8} and $(\pmb{p}^u, \pmb{\zeta}^u,\pmb{\nu}^u, \pmb{\alpha}^u, \pmb{\beta}^u) $ represent the unbounded directions. 

We use the lift-and-project cutting-plane method for solving the binary optimization in \eqref{eq9}. For completeness, we describe this method in Appendix \ref{app:2}.  The final algorithm is summarized in Algorithm \ref{alg:2}.

\begin{algorithm}[!htb]
	\SetAlgoLined
	\KwResult{ Frequency setting for MT, rebalancing flows for MoD, and prices. }
	Initialization:  $\pmb{z} = 0$ for all $z_{\ell,f}$ and $z_{\psi}^{b_{ij}}$\;
	\For{ $\ell \in [L] $}{
		Update $Z_{\ell} = [z_{\ell, f}]_{f\in [F]}$ such that, if $z_{\ell, f}>0$, set $z_{\ell, f'}=1$ for all $f' < f$ and $z_{\ell, f'}=0$ for all  $f'>f$.
		\For{ $f \in [F]$ }{
			Solve the LP-relaxation of \eqref{eq9} to obtain $Z^*$\;
			Generate lift-and-project cuts with $Z^*$ for $Z\in \mathbb{R}^{L\times F}$ for \eqref{eq9} (Appendix \ref{app:1}) \;
			Update all bundled services $z_{\psi}^{b_{ij}}$.
			}
		Solve subproblem \eqref{eq5} by Algorithm \ref{alg:1} and generate either feasibility or optimality cuts\;
	}
	\caption{Algorithm for joint pricing and transit network design problem} \label{alg:2}
\end{algorithm}




\section{Numerical experiment} \label{sec:4}
\subsection{Experiment setup and data description}
We conduct a numerical experiment that retrofits the current MTA bus system in Nashville, Tennessee with MoD as a first or last-mile connection. Residents in Nashville, like most metropolitan areas in the United States, have unequal access to transit systems. Therefore, determining the efficacy of extending the MT structure with the assistance of MoD connections is a meaningful activity.

Table \ref{table:1} summarizes the data used in the experiment. The trip demand data estimates the origin-destination matrix at the census tract level of daily commuting trips, which is used for $\lambda_{st}$. There are two types of travelers $\theta\in \{1,2 \}$ corresponding to price-sensitive and QoS-sensitive travelers.
We can include more types of travelers by expanding set $\Theta$ in other applications. 
For all o-d pairs, $q_{st}(1) = q_{st}(2) = 0.5$. For each kind of traveler, the utility functions follow
\begin{align}
	v_{ \theta m} =  \begin{cases}
		p_{\theta m}^2 + \sum_{e_{ij} \in E_m} (c_f \cdot f_e - c_l \cdot l_e) & \theta = 1 \\
		p_{\theta m} + \sum_{e_{ij} \in E_m} ( c_f \cdot f_e^2 - c_l \cdot l_e)  & \theta = 2
	\end{cases}.
	\nonumber
\end{align}  

Each candidate mode set $\mathcal{M}_{st}$ includes all combinations of bus routes with no more than two transfers and MoD as first- or last-mile connections.  We truncated $\mathcal{M}_{st}$ in long-distance trips if there exist more than five hybrid combinations of MoD and MT, because travelers are not capable of evaluating a large size of alternative options. For each combination, travelers consider all possible frequencies of MT as in Proposition \ref{prop1}.

\begin{table}[!htb]
	\caption{Summary of numerical experiment data sources } \label{table:1}
	\begin{tabular}{l|l|l}
		\toprule
		Item            & Datasource & Description \\ \hline
		Trip demand            &  Access Nashville \cite{nashville2020}      &     \begin{tabular}[c]{@{}l@{}} Origin-destination   \\ movement matrix  \end{tabular}           \\ \hline
		Bus route            &   Nashville MTA  \cite{nashvilleMTA2020}       &    GIS and GTFS         \\ \hline
		\begin{tabular}[c]{@{}l@{}}  Cost  \\ parameters  \end{tabular}     &  Operational cost \cite{banerjee2020market}          &    \begin{tabular}[c]{@{}l@{}} MT cycling time,   \\ setup cost, \\ Rebalancing \\ cost of MoD \end{tabular}                 \\ 
		\bottomrule
	\end{tabular}

\end{table}

\begin{figure}[!htb]
	\centering
	\includegraphics*[width = 0.48\textwidth]{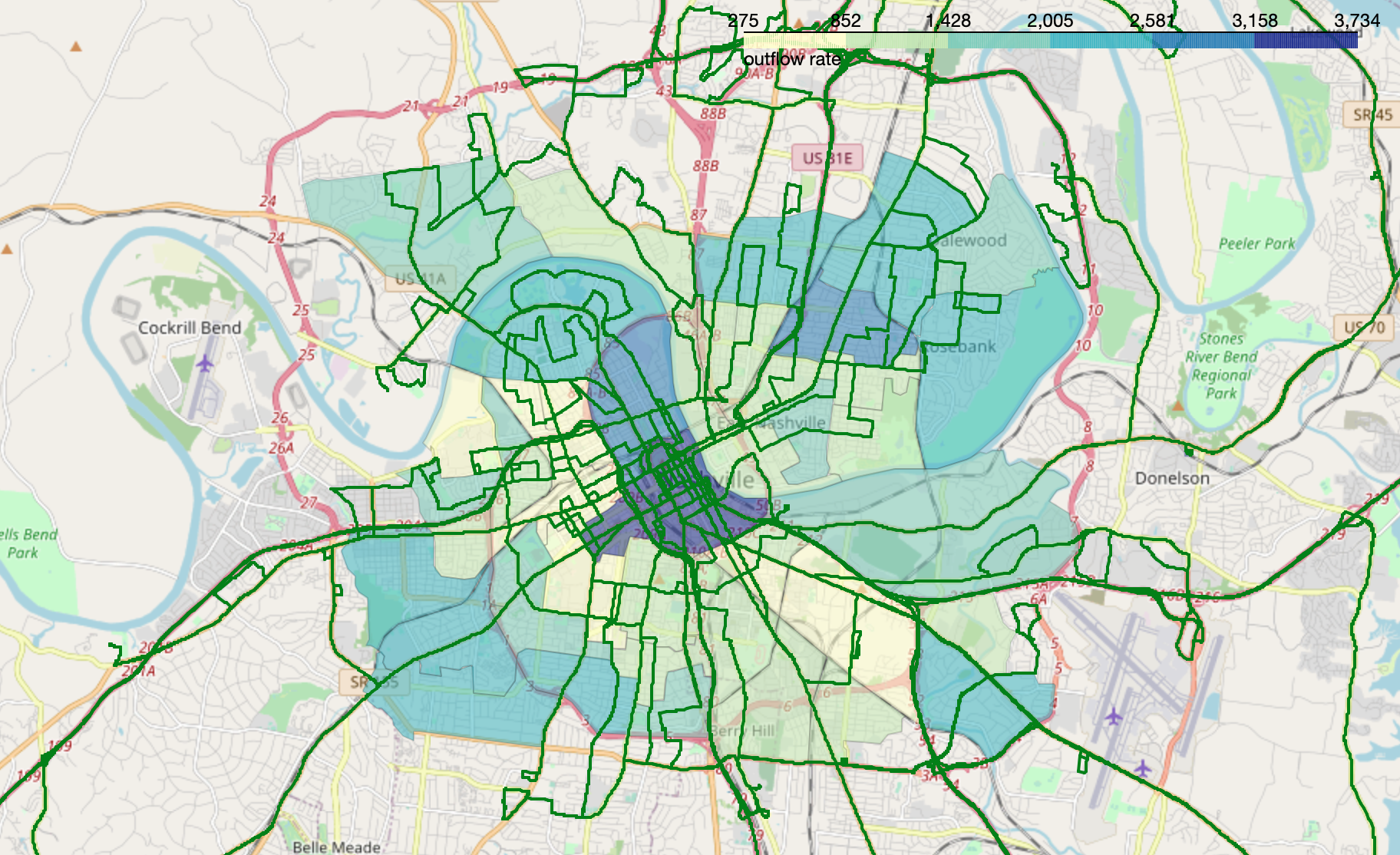}
	\caption{ Nashville MTA bus system (green lines) and origin-destination flow (outflow) data (chloropleth map)   } \label{fig:4}
\end{figure}

The MTA bus routes and trip demand data are shown in Figure \ref{fig:4}.  The size of optimization in \eqref{eq3} depends on fixed network properties such as the scale of MT network and the dimension of trip demand od matrix, as well as user-specific design parameters such as the MT frequency levels $F$ or the total number of MT vehicles $\mathcal{R}$. Table \ref{table:2} summarizes the benchmark model used throughout the remaining analysis.

\begin{table}[!htb]
	\caption{Summary of parameters in Nashville MTA case study} \label{table:2}
	\begin{tabular}{l|c|c}
		\toprule
		Item & Notation & Value \\ \hline
		Bus lines &     $L$     &  $48 $    \\ \hline
		Bus frequency level &      $F$    & $5 $     \\ \hline
		  \begin{tabular}[c]{@{}l@{}}  Number of vertices   \\  (census tracts) \end{tabular}        &   $|V|$       &   $47$    \\ \hline
		 \begin{tabular}[c]{@{}l@{}}  Number of o-d pairs  with \\ $\lambda_{st} >0$   \end{tabular}      & $(s,t)$ & $836$ \\ \hline 
		Number of modes &  $|\mathcal{M}_{st} |$       &  $1$ to $5$     \\ \hline
		 \begin{tabular}[c]{@{}l@{}} Number of MT frequency  setting    \\   variables   \end{tabular}   & $z_{\ell, f}$ & $240$ \\ \hline
		 Number of bundled variable &  $z_{\psi}^{b_{ij}}$  &  $320$ \\ \hline 
		 Number of choice variables & $x_{\theta m}$ & $ 139,744$ \\ \hline
		 \begin{tabular}[c]{@{}l@{}} Number of MoD rebalancing   \\ variables   \end{tabular}    & $r_{ij}$ & $664$  \\ \hline
		 Total number of buses & $\mathcal{R}$ & $20$ \\ \hline
		 Vehicle capacity & $\mathcal{V}_{\ell}$ & $75$  \\
		\bottomrule
	\end{tabular}
\end{table}

\subsection{Optimal prices and transit network design in Nashville, TN}
\begin{figure}[!htb]
	\centering
	\includegraphics*[width = 0.48\textwidth]{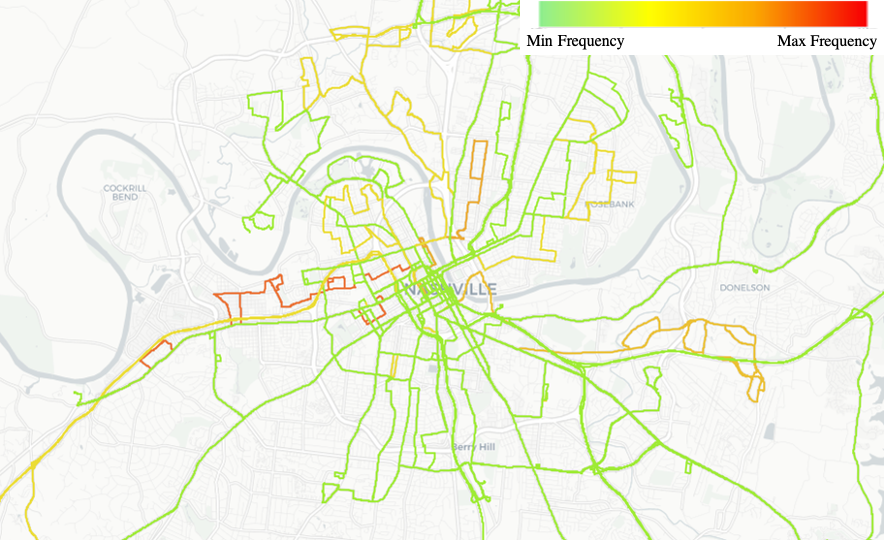}
	\caption{ Optimal frequency setting for Nashville MTA bus system    } \label{fig:5}
\end{figure}

The optimal frequency setting of the MTA bus system is shown in Figure \ref{fig:5}. The  lines connecting  suburban regions to the downtown area are expected to run more frequently than the downtown area. MoD carries travelers to these lines and aggregate flow via these main corridors. This result, on the other hand, has limitations since o-d trip demand data predicts everyday commuting trips predominantly. Furthermore, the proportion of the population able to engage in this multimodal mobility service is not specified. 

The computational efficiency of the proposed algorithm is shown in Table \ref{table:3}. Two benchmarks that we compared with are state-of-the-art MIP solvers (Gurobi 9.0 on a 1.1 GHz Dual-Core processor). The runtime of the proposed algorithm is significantly shorter for both cases. When the size of the network or the frequency level increases, our approach outperforms the MIP solver as the proposed algorithm only solves a sequence of LP.   Besides, the MIP solvers cannot return the dual prices directly because it is not well-defined. In contrast, as the subproblems are multicommodity flow problems with rebalancing constraints, dual prices are well-defined for each $(s,t)$ path in the multimodal network.

\begin{table}[!htb]
	\caption{Computational results for the Nashville MTA case study} \label{table:3}
	\begin{tabular}{l|c|c|c|c|c|c}
		\toprule
		\multicolumn{1}{c|}{\multirow{2}{*}{ {\small Method  } }}                                             & \multicolumn{2}{c|}{ \begin{tabular}[c]{@{}l@{}} {\small  Runtime  }   \\ {\small  (second) }   \end{tabular}      } & \multicolumn{2}{c|}{ \begin{tabular}[c]{@{}l@{}} UB-LB   \\  gap (\%)  \end{tabular}  }  &  \multicolumn{2}{c}{ {\small Objective }   }  \\ \cline{2-7} 
		\multicolumn{1}{c|}{}                                                                    & {\scriptsize  Inst. 1   }      &  {\scriptsize Inst. 2   }    & {\scriptsize Inst. 1   }       &  {\scriptsize Inst. 2  }    & {\scriptsize Inst. 1 }   & {\scriptsize Inst. 2  }     \\ \hline
		{\scriptsize  \begin{tabular}[c]{@{}l@{}}MIP solver with  \\ conservative cut \\  generation\end{tabular}   }&       $3600^*$       &     $2832$             &        $20.74\%$            &      $0.0\%$  & $4246$   &   $5932$        \\ \hline
		{\scriptsize \begin{tabular}[c]{@{}l@{}}MIP solver with \\ automatic cut \\ generation\end{tabular}   }   &      $ 3600^*$          &    $2422 $             &     $ 21.93\%$      &        $0.0\%$    & $4050$   & $5569$    \\ \hline
	  {\scriptsize \begin{tabular}[c]{@{}l@{}}Benders decom- \\
	  		position and \\ approximation \\ algorithm \end{tabular}   }	      &      $1232$           &      $  921     $      &             -     &          -   &  $4815$   &  $5208$      \\ \bottomrule
	\end{tabular}
	{\scriptsize
	\begin{itemize}
		\item Instance 1 is from Table \ref{table:2}.
		\item 	Instance 2 changes the frequency level $F = 3$ and types of customers $|\Theta| =3$. 
		\item  Runtime with $*$ means the solver hits the time limits and stops early. 
		\item $UB-LB$ gap is the difference between the upper bound and lower bound of the solutions.
	\end{itemize}
}
\end{table}

\subsection{Sensitivity analysis}
\begin{figure}[!htb]
	\subfloat[Sensitivity on total vehicle number]{  \includegraphics[ width= 0.4\textwidth ]{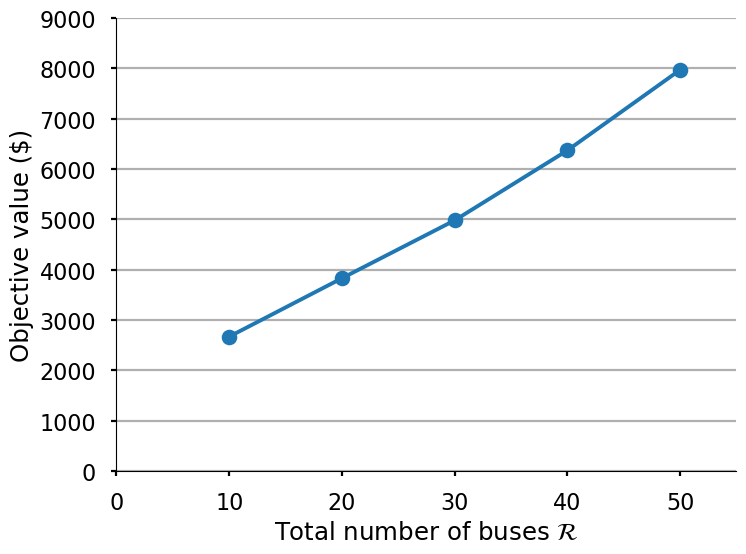} } \\
	\subfloat[Sensitivity on vehicle capacity]{  \includegraphics[ width= 0.4\textwidth ]{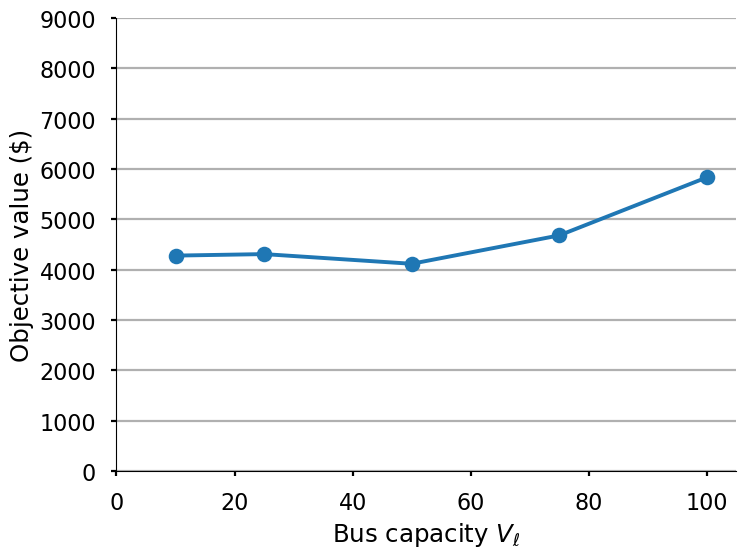} } \\
	\caption{Sensitivity analysis of the optimal value regarding MT parameters } \label{fig:6}
\end{figure}

Since several operational parameters of the status-quo system are unknown, we conduct a sensitivity analysis in Figure \ref{fig:6}. The most critical design parameters for MT are the total number of buses $\mathcal{R}$ and vehicular capacities $V_{\ell}$. We test the optimal design's sensitivity with regard to these two parameters.

As the MT system's available resources  increase, the overall welfare increases significantly as more travelers can save money by using MT facilities. Note that we do not consider the purchase costs of additional MT buses, but assume the current system's bus fleet size may vary from Table \ref{table:2}.  On the other hand, it is not inherently a profitably practice to use  large-capacity vehicles as we can raise the frequencies on specific routes to offset the reduced  per-vehicle capacity. This coincides with the recent development in operating small-size flexible MT.   

\section{Conclusion}  \label{sec:5}
As a building block for any multimodal mobility ecosystem, this work studies a joint pricing and transit network design problem. A unified MILP framework solves the frequency setting of MT in the master problem and solves the price and MoD rebalancing in subproblems sequentially. This primal-dual approach is substantially easier to implement and more efficient in computation compared to the traditional bilevel program with equilibrium constraints. With the advent of a shared, on-demand, and diverse mobility economy, a case study in Nashville, TN, highlights the ability of using this approach in solving real-world challenges.

The limitations of this work include the type of choice model for travelers that can be accommodated by this linear model, even though it admits the MNL, which is the most common model of discrete choice in transportation. Furthermore, we presume that information in the system is publicized such that each traveler is completely rational in making their travel plans, which might not be the case in reality. We leave these questions for future work.

\section*{Acknoledgement}  \label{sec:6}
This material is based upon work partially supported by the National Science Foundation under Grant No. CNS-1952011 and CMS-1839346.


\bibliographystyle{ACM-Reference-Format}
\bibliography{sample-sigplan}

\newpage 

\appendix

\section{Summary of notation}  \label{app:1}
\noindent \begin{table}[!htb]
\caption{Summary of notation} \label{table:4}
\begin{tabular}{c|l}
\toprule
Notation & Definition \\ \hline
\multicolumn{2}{l}{Decision variables:} \\ \hline
     $x_{\theta m}$    & \begin{tabular}[c]{@{}l@{}} Type $\theta$ traveler's mode choice strategy \\ for mode $m$      \end{tabular}      \\ \hline
      $z_{\ell, f}$   &  \begin{tabular}[c]{@{}l@{}} Binary variable for setting MT route $\ell$'s  \\ frequency at level $f$  \end{tabular}    \\ \hline
     $z_{\psi}^{b_{ij}} $    &  \begin{tabular}[c]{@{}l@{}} Binary variable for setting bundled MT\\  services  $b_{ij}$'s frequency at level $\psi$  \end{tabular}       \\ \hline
     $r_{ij}$    &  MoD's rebalancing flows on edge $e_{ij}$       \\ \hline
     $y_{\theta m }$    & Intermediate variable $y_{\theta m } = \lambda_{st}(\theta) x_{\theta m}$           \\ \hline
     $p_{\theta m}$ &  \begin{tabular}[c]{@{}l@{}}  Prices for mode $m$ charged for type  $\theta$ \\  travelers \end{tabular}    \\ \hline 
     $\pmb{p, \zeta, v, \alpha, \beta}$ & Dual variables of subproblem     \\ \hline
     $\tau, \phi$ &   \begin{tabular}[c]{@{}l@{}} Intermediate variables for benders' \\ decomposition \end{tabular}        \\ \midrule 
     \multicolumn{2}{l}{Parameters:} \\ \midrule 
      $G = (V,E)$    &  \begin{tabular}[c]{@{}l@{}}  Graph of multimodal mobility network   \\ consists of vertices $V$ and edges $E$ \end{tabular}         \\ \hline
      $n$ &  Cardinality of vertices $V$ \\ \hline 
      $\Theta$ & Type space of travelers \\ \hline 
     $\lambda_{st}$    & Travel demand from vertex $s$ to vertex $t$   \\ \hline
      $q_{st}(\theta) $   &  \begin{tabular}[c]{@{}l@{}}  Probability density of type $\theta$ for \\ $s-t$ trip \end{tabular}           \\ \hline
      $L$   & Total number of MT lines          \\ \hline
      $F$   & Maximum frequency level for each line    \\ \hline
      $\mathcal{M}_{st}$   & \begin{tabular}[c]{@{}l@{}}  Set of hybrid modes available for \\ $s-t$ trip \end{tabular} \\ \hline 
      $v_{\theta m}$  & Valuation of mode $m$ to type $\theta$ travelers \\  \hline
       $E_m$  &  Edges trespassed by mode $m$          \\ \hline
      $V_m$  &  Vertices trespassed by mode $m$          \\ \hline
      $\mathcal{B}_{ij}$  & Set of bundled MT services on edge $e_{ij}$           \\ \hline
      $o_{\ell}$    &  Unit setup cost for line $\ell$          \\ \hline
      $c_m$     & Operational cost of mode $m$           \\ \hline
      $\mathcal{R}$      & Total number of available MT vehicles            \\ \hline
      $T_{\ell}$    & Cycling time of line $\ell$           \\ \hline
      $\mathcal{P}^{st}_{i}$ & \begin{tabular}[c]{@{}l@{}}  Set of paths for $i^{th}$ commodity's  $s-t$   \\ trip in  subproblem's approximation \\ algorithm   \end{tabular}          \\ \hline
      $K$  & Number of commodities in subproblem \\ \hline 
      $\epsilon, \delta$         & \begin{tabular}[c]{@{}l@{}} Error tolerance in approximation \\ algorithm \end{tabular}       \\ \hline
      $\mathcal{V}_{\ell}$  & Vehicle capacity on line $\ell$           \\  \midrule
      \multicolumn{2}{l}{Acronyms:} \\ \midrule 
      MoD & Mobility-on-Demand\\ \hline
      MNL &  Multinomial Logit model \\
      \bottomrule
\end{tabular}
\end{table}

\section{Lift-and-project cuts for master problem} \label{app:2}
We first apply the convexification procedure to all constraints in \eqref{eq9}. 
For example, we can denote the left-hand coefficient matrix of \eqref{eq3b}  as $A_2$. Since $z_{\ell,f} \geq 0$, we can rewrite the constraint as
\begin{align}
	z_{\ell, f} (Z^{b_{ij} }  - A_2^{b_{ij}} Z_{\ell'}  ) \leq 0, \qquad  \forall b_{ij}\in \mathcal{B}_{ij} , \forall e_{ij} \in E  \nonumber \\
	(1-  z_{\ell, f} ) (Z^{b_{ij} }  - A_2^{b_{ij}} Z_{\ell'}  ) \leq 0, \qquad  \forall b_{ij}\in \mathcal{B}_{ij} , \forall e_{ij} \in E  \nonumber
\end{align}
and linearize the constraints as
\begin{align}
	\begin{cases}
		z_{\ell, f}^2 \leftarrow z_{\ell, f}  \\ 
		z_{\ell, f} z_{\ell', f'} \leftarrow 	w_{\ell, f}  , \qquad \ell \neq \ell \text{ or } f' \neq f
	\end{cases}. \nonumber 
\end{align}

Denote the generic form of constraints in  the LP-relaxation as $A Z \leq b$. Given the fractional solution $Z^*$, the lift-and-project cuts are generated by the following cut generating LP:
\begin{align}
	& \text{maximize }  \kappa Z^*  - \kappa_0 \\
	s.t. \, &  \kappa \leq u A + u_0 \pmb{e}_{\ell, f} \nonumber  \\
	& \kappa \leq u A -  v_0 \pmb{e}_{\ell, f} \nonumber \\ 
	& \kappa_0 \geq u b  \nonumber  \\
	& \kappa_0 \geq v b - v_0 \nonumber  \\
	& \pmb{1}^{\intercal} u + u_0 + \pmb{1}^{\intercal} v + v_0 = 1  \nonumber \\
	& u, v, u_0, v_0 \geq 0 \nonumber 
\end{align}
where $\pmb{e}_{\ell, f}$ is an all-zero vector but one at the $\ell, f$ entry. 

Let  $A_{\ell, f}$ be the  column index of $A$. We add the strengthened lift-and-project cuts for $z_{\ell', f'}$ as follows:
\begin{align*}
	& \sum_{\ell \in [L], f\in [F]}  \alpha_{ \ell, f } z_{\ell, f} \leq \beta, \\
	&	\text{where } \alpha_{\ell, f} =  \\
	& \begin{cases}
		\max\{  u^{\intercal} A_{\ell, f} + u_0 \lceil m_{\ell,f} \rceil,  v^{ \intercal } A_{\ell,f} - v_0 \lfloor m_{\ell, f} \rfloor   \}, \, (\ell', f') \neq (\ell, f) \\
		\max\{  u^{\intercal} A_{\ell, f} - u_0,  v^{ \intercal } A_{\ell, f} + v_0  \}, \quad (\ell', f') = (\ell, f) \\
		\max \{u^{\intercal} A_{\ell, f},  v^{ \intercal } A_{\ell, f}  \}, \quad  \ell = L, f = F
	\end{cases} \\
	& \beta = \min\{ u^{\intercal} b, v^{\intercal} b + u_0  \} \\
	& m_{\ell, f} = \frac{ v^{\intercal } A_{\ell, f} - u^{\intercal} A_{\ell, f} }{ u_0 + v_0 }.
\end{align*}

\end{document}